\documentclass[reqno,11pt]{amsart}
\usepackage{multind}
\usepackage{hyperref}
\usepackage{graphicx}
\usepackage{amscd}
\usepackage{amssymb}
\usepackage[mathscr]{eucal}
\numberwithin{equation}{section}
\allowdisplaybreaks[1]

\newtheorem{Def}{Definition}[section]
\newtheorem{Thm}[Def]{Theorem}

\newcommand{\beq}{\begin{equation}}
\newcommand{\eeq}{\end{equation}}
\newcommand{\Proof}{\begin{proof}}
\newcommand{\QED}{\end{proof} \noindent}

\newcommand{\spc}{\;\;\;\;\;\;\;\;\;\;}

\newcommand{\bra}{\,<\!\!}
\newcommand{\ket}{\!\!>\,}
\newcommand{\C}{\mathbb{C}}
\newcommand{\R}{\mathbb{R}}
\newcommand{\1}{\mbox{\rm 1 \hspace{-1.05 em} 1}}

\renewcommand{\H}{\mathscr{H}}
\newcommand{\F}{\mathscr{F}}

\newcommand{\psiu}{\psi^\uparrow}
\newcommand{\psid}{\psi^\downarrow}

\newcommand{\Sl}{\mbox{$\prec \!\!$ \nolinebreak}}
\newcommand{\Sr}{\mbox{\nolinebreak $\succ$}}
\newcommand{\SU}{{\rm{SU}}}

\newcommand{\Og}{{\mathcal{O}}}
\begin{document}
\title[The Fermionic Projector, Entanglement and Collapse]{The Fermionic Projector, Entanglement, \\
and the Collapse of the Wave Function}

\author[F.\ Finster]{Felix Finster \\ \\ November 2010}
\thanks{Supported in part by the Deutsche Forschungsgemeinschaft.}
\address{Fakult\"at f\"ur Mathematik \\ Universit\"at Regensburg \\ D-93040 Regensburg \\ Germany}
\email{Felix.Finster@mathematik.uni-regensburg.de}

\maketitle

\begin{abstract}
After a brief introduction to the fermionic projector approach,
we review how entanglement and second quantized bosonic
and fermionic fields can be described in this framework.
The constructions are discussed with regard to decoherence 
phenomena and the measurement problem.
We propose a mechanism leading to the collapse of the
wave function in the  quantum mechanical measurement process.
\end{abstract}

\tableofcontents

\section{An Introduction to the Fermionic Projector Approach} \label{secPFP}
In this section we give a brief introduction to the framework of the fermionic projector,
outlining a few ideas, methods and results (for details see~\cite{PFP, sector}
and the references in the review papers~\cite{lrev, srev}).

\subsection{The Idea of Spontaneous Structure Formation}
Due to the ultraviolet divergences of quantum field theory and the difficulties in quantizing
gravity, it is generally believed that our notions of space and time should be modified on
a microscopic scale, usually thought of as the Planck scale.
However, there is no agreement on what the right framework for physics on the Planck scale
should be. A major difficulty is that if one modifies the microscopic structure of Minkowski
space or of a Lorentzian manifold in a naive way, the gauge invariance and the diffeomorphism invariance
of the resulting physical theory are not preserved. For example, a simple
ultraviolet regularization is to replace Minkowski space by a four-dimensional lattice.
If one now works instead of partial derivatives with difference quotients,
then the gauge invariance is lost. A better method is to work with the integrals of the gauge
potentials along the links of the lattice, which restores gauge invariance and leads to the framework
of lattice gauge theories. However, lattice gauge theory is not compatible with the equivalence principle,
because after performing a diffeomorphism, the lattice points are no longer in a regular configuration.
Taking a somewhat different point of view, the problem of a lattice is that it uses
structures (like the nearest neighbor relation or the lattice spacing) which refer to
an ambient space-time, although we wanted to avoid this background space-time
for the formulation of the physical model.
More generally, one would like to have a framework which is background independent
and respects the fundamental physical principles.

The fermionic projector approach provides a framework which meets the above requirements.
We begin with a finite set of space-time points~$M=\{1, \ldots, m\}$, without presupposing any
relations between the space-time points. On this finite set of points, we then define an ``ensemble
of wave functions''~$\psi_1, \ldots, \psi_f$. We introduce an action~$\mathcal{S}$
which to every configuration of wave functions associates a positive number.
By minimizing~$\mathcal{S}$, the wave functions tend to get into a configuration where the action is
smaller. This process can be understood as a self-organization of the wave functions, driven by
our action principle. In the resulting minimizing configuration, the wave functions are typically
delocalized in the sense that they are non-zero on many space-time points.
Thus by considering the correlations of the wave functions
at different space-time points, one gets relations between these points.
With this mechanism, referred to as {\em{spontaneous structure formation}},
additional structures and objects arise in space-time. The idea is that all the structures needed for
the formulation of physics, like the causal structure, gauge potentials and fields, the Lorentzian
metric, etc., are generated in this way. Finally, evaluating the Euler-Lagrange equations
corresponding to our action principle in terms of all these structures should give
the physical equations.

Before we can make this idea mathematically precise, we need to specify in which space
the wave functions~$\psi_i$ should take their value. To this end, we associate to every space-time
point~$x \in M$ a complex vector space~$S_x$, referred to as the {\em{spin space}}.
We endow this vector space with an indefinite inner product~$\Sl .|. \Sr_x$, the {\em{spin scalar product}}.
Then a wave functions at~$x$ takes values in the corresponding spin space,
\[ \psi_i \::\: x \mapsto \psi_i(x) \in S_x\:. \]
In physical applications, a wave function should correspond to a four-component Dirac
wave function with the usual inner product~$\Sl \psi | \phi \Sr_x = \overline{\psi(x)} \phi(x)$,
where~$\overline{\psi} = \psi^\dagger \gamma^0$ is the adjoint spinor
(this will be made precise in Section~\ref{secminkowski} below). 
With this in mind, we assume that~$S_x$ is four-dimensional, and that the spin scalar product
is indefinite of signature~$(2,2)$. Taking the direct sum of all the spin spaces,
\[ H = \bigoplus_{x \in M} S_x\:, \qquad
\bra \psi | \phi \ket := \sum_{x \in M} \Sl \psi(x) | \phi(x) \Sr_x \:, \]
we obtain an inner product space~$(H, \bra .|. \ket)$ of dimension~$4m$ and signature~$(2m, 2m)$.
We can then describe the evaluation of a wave function at a space-time point~$x$
by a projection operator~$E_x$,
\[ \psi(x) = E_x \,\psi \qquad \text{with} \qquad E_x \::\: H \rightarrow S_x \::\: \psi \mapsto \psi(x)\:. \]
Moreover, it is convenient to describe the ensemble of wave functions~$\psi_1, \ldots, \psi_f \in H$ by
a projector~$P$ on the subspace of~$H$ spanned by these wave functions.
This motivates the framework of fermion systems in discrete space-time which we now introduce.

\subsection{An Action Principle for Fermion Systems in Discrete Space-Time} \label{secdiscrete}
We let~$(H, \bra .|. \ket)$ be a finite-dimensional complex vector space endowed with an indefinite
inner product (thus~$\bra .|. \ket$ is non-degenerate, but not positive definite).
Next, we let~$M = \{1, \ldots, m\}$ be a finite set. To every point~$x \in M$ we associate a
projector $E_x$ (a projector in~$H$ is defined just as in Hilbert spaces as a linear
operator which is idempotent and symmetric). We assume that these projectors are orthogonal and
complete in the sense that
\beq \label{oc}
E_x\:E_y \;=\; \delta_{xy}\:E_x \spc {\mbox{and}} \spc
\sum_{x \in M} E_x \;=\; \1\:.
\eeq
Furthermore, we assume that the images~$E_x(H) \subset H$ of these
projectors are non-de\-ge\-ne\-rate subspaces of~$H$, which
all have the same signature~$(2,2)$.
The points~$x \in M$ are called {\em{discrete space-time points}}, and the corresponding
projectors~$E_x$ are the {\em{space-time projectors}}. The
structure $(H, \bra .|. \ket, (E_x)_{x \in M})$ is called {\em{discrete space-time}}.
The particles of our system are described by one
more projector~$P$ in~$H$, the so-called {\em{fermionic projector}}, which
has the property that its image~$P(H)$ is a
negative definite subspace of~$H$.
The resulting system~$(H, \bra .|. \ket, (E_x)_{x \in M}, P)$ is referred to as
a {\em{fermion system in discrete space-time}}.

Let us briefly discuss these definitions and introduce a convenient notation.
The vectors in the image of~$P$ have the interpretation as the
occupied fermionic states of our system, and thus the rank of~$P$
gives the {\em{number of particles}} $f := \dim P(H)$.
The space-time projectors~$E_x$ can be used to project vectors of~$H$
to the subspace~$E_x(H) \subset H$. Using a more graphic notion, we
also refer to this projection as the {\em{localization}} at the space-time point~$x$.
For the localization of a vector~$\psi \in \H$ we use the short notation
\beq \label{wave}
\psi(x) := E_x\,\psi
\eeq
and refer to~$\psi(x)$ as the corresponding {\em{wave function}}.
The wave functions at~$x$ span the vector space~$E_x(H)$, referred to as the {\em{spin space}}
at~$x$ and denoted by~$S_x$. It is endowed with the inner product~$\bra .| E_x . \ket$
of signature~$(2,2)$, which we also denote by~$\Sl .|. \Sr$ and refer to as the
{\em{spin scalar product}}. Using the relations~\eqref{oc},
we can then write
\beq \label{stipdiscrete}
\bra \psi | \phi \ket = \sum_{x \in M} \Sl \psi(x) | \phi(x) \Sr \:.
\eeq
The localization of the fermionic projector is denoted by~$P(x,y) := E_x\,P\,E_y$.
This operator maps~$S_y\subset H$ to~$S_x$, and we usually
regard it as a mapping between these subspaces,
\[ P(x,y) = E_x\,P\,E_y \::\: S_y \rightarrow S_x \:. \]
Again using~\eqref{oc}, we can write the wave function corresponding to~$P \psi$ as follows,
\[ (P\psi)(x) \;=\; E_x\: P \psi \;=\; \sum_{y \in M} E_x\,P\,E_y\:\psi
\;=\; \sum_{y \in M} (E_x\,P\,E_y)\:(E_y\,\psi) \:, \]
and thus
\beq \label{diskernel}
(P\psi)(x) \;=\; \sum_{y \in M} P(x,y)\: \psi(y)\:.
\eeq
This relation resembles the representation of an operator with an integral kernel, and therefore we call~$P(x,y)$ the {\em{discrete kernel}} of the fermionic projector.

In order to introduce our action principle, for any~$x, y \in M$
we define the {\em{closed chain}}~$A_{xy}$ by
\beq \label{cc}
A_{xy} \;=\; P(x,y)\: P(y,x) \::\: S_x \rightarrow S_x \:.
\eeq
We denote the eigenvalues of~$A_{xy}$ counted with algebraic multiplicities
by~$\lambda^{xy}_1,\ldots,\lambda^{xy}_{4}$. We define the Lagrangian~${\mathcal{L}}$ by
\beq \label{Lcrit}
{\mathcal{L}}[A_{xy}] = 
\sum_{i,j=1}^4 \left( |\lambda^{xy}_i| - |\lambda^{xy}_j| \right)^2
\eeq
(for simplicity we here leave out the constraints and consider only the so-called critical case of the
auxiliary Lagrangian). Our action principle is to
\beq
\text{minimize the action} \quad {\mathcal{S}}[P] = \frac{1}{8} \sum_{x,y \in M}
{\mathcal{L}}[A_{xy}] \:, \label{action}
\eeq
where we consider variations of the fermionic projector, keeping the number of particles
as well as discrete space-time fixed. For a more general mathematical exposition of this
variational principle and the existence theory we refer to~\cite{discrete}.
For a discussion of the underlying physical principles see~\cite[Section~2]{rev}.
We finally remark that generalizing the above setting leads to the framework 
of {\em{causal fermion systems}} (see~\cite{rrev} and the references therein).

\subsection{Effects of Spontaneous Structure Formation}
We now discuss two effects of spontaneous structure formation:
the spontaneous breaking of the permutation symmetry and the generation of a discrete
causal structure.

We first point out that in the above definitions, we did not
distinguish an ordering of the space-time points; all our notions are symmetric under permutations
of the points of~$M$. This suggests that a minimizing fermionic projector should also be
permutation symmetric. Surprisingly, this is not the case.
Indeed, in~\cite{osymm} it is shown under general assumptions that that the fermionic projector
cannot be permutation symmetric. This shows rigorously that the effect of spontaneous
structure formation really occurs, giving rise to a non-trivial symmetry group acting on the
discrete space-time points.
Before stating the main result, we need to specify what we mean by a
symmetry of our fermion system. The group of all permutations of the space-time points is the
symmetric group, denoted by~$S_m$.
\begin{Def} \label{defouter}
A subgroup~$\Og \subset S_m$
is called {\bf{outer symmetry group}} of the fermion system in discrete space-time if
for every~$\sigma \in \Og$ there is a unitary transformation~$U$ such that
\beq \label{USdef}
UPU^{-1} \;=\; P \qquad {\mbox{and}} \qquad
U E_x U^{-1} \;=\; E_{\sigma(x)} \quad
{\mbox{for all $x \in M$}}\:.
\eeq
\end{Def}

\begin{Thm} {\bf{(spontaneous breaking of the  permutation symmetry)}} \label{thmend}
Suppose that $(H, \bra .|. \ket, (E_x)_{x \in M}, P)$ is a
fermion system in discrete space-time. Assume that the number of space-time points
and the number of particles~$f$ are in the range
\[ m > 10 \qquad \text{and} \qquad 2 < f < m-1 \:. \]
Then the fermion system cannot have the outer symmetry group~$\Og = S_m$.
\end{Thm} \noindent
In particular, this theorem applies in the physically interesting case of many particles and
even more space-time points.

More detailed information on the space-time relations generated by the spontaneous structure
formation is obtained by analyzing the eigenvalues of the closed chain.
Note that in an indefinite inner product space, the eigenvalues of a self-adjoint operator~$A$ need
not be real, but alternatively they can form complex conjugate pairs (see~\cite{GLR}
or~\cite[Section~3]{discrete}). This fact can be used to introduce a notion of causality.

\begin{Def} {\bf{(causal structure)}} \label{defcausal}
Two space-time points~$x,y \in M$ are called {\bf{timelike}} separated if
the spectrum of~$A_{xy}$ is real. The points are
{\bf{spacelike}} separated if the spectrum of~$A_{xy}$ forms two complex
conjugate pairs having the same absolute value. In all other cases, the two
points are {\bf{lightlike}} separated.
\end{Def} \noindent
This notion is compatible with our action principle in the following sense.
Suppose that two space-time points~$x$ and~$y$ are spacelike separated.
Then the eigenvalues~$\lambda^{xy}_i$ all have the same absolute value,
and thus the Lagrangian~\eqref{Lcrit} vanishes.
A short calculation shows that the first variations of the action
also vanish, so that~$A_{xy}$ does not enter the corresponding Euler-Lagrange
equations. This can be seen in analogy to the usual notion of causality
that points with spacelike separation cannot influence each other.

We point out that by analyzing the kernel of the fermionic projector in more detail,
one can even distinguish a direction of time, define a parallel transport of spinors and introduce
curvature, thus giving a proposal for a ``Lorentzian quantum geometry.'' The interested reader
is referred to~\cite{lqg} and the survey article~\cite{rrev}.

\subsection{The Correspondence to Minkowski Space} \label{secminkowski}
We now describe how to get a connection between discrete space-time and Minkowski space.
The simplest method for getting a correspondence to relativistic quantum mechanics in
Minkowski space is to replace the discrete space-time points~$M$ by the space-time
continuum~$\R^4$ and the sums over~$M$ by space-time integrals. For a
vector~$\psi \in H$, the corresponding localization~$E_x \psi$
should be a 4-component Dirac wave function, and the spin scalar product $\Sl \psi | \phi \Sr$
on~$E_x(H)$ should correspond to the usual Lorentz invariant scalar product on Dirac
spinors~$\overline{\psi} \phi$
with~$\overline{\psi} = \psi^\dagger \gamma^0$ the adjoint spinor.
In view of~\eqref{diskernel}, the discrete kernel should go over to the
integral kernel of an operator~$P$ on the Dirac wave functions,
\[ (P \psi)(x) \;=\; \int_M P(x,y)\, \psi(y)\: d^4y \:. \]
The image of~$P$ should be spanned by the occupied fermionic states. We take Dirac's
concept literally that in the vacuum all negative-energy states are occupied by fermions
forming the so-called {\em{Dirac sea}}. This leads us to describe the vacuum by the
integral over the lower mass shell
\beq \label{Psea}
P^\text{sea}(x,y) \;=\; \int \frac{d^4k}{(2 \pi)^4}\: (k_j \gamma^j+m)\:
\delta(k^2-m^2)\: \Theta(-k^0)\: e^{-ik(x-y)} \:.
\eeq
(Here~$\Theta$ is the Heaviside function, and in the terms~$k^2$ and~$k(x-y)$
we use the Minkowski inner product, using the signature convention~$(+ - - - )$).
Computing the Fourier integral, one sees that~$P(x,y)$ is a smooth function,
except on the light cone~$\{ (y-x)^2=0\}$, where it has poles and singularities.

Let us compare Definition~\ref{defcausal} with the usual notion of causality
in Minkowski space. Even without computing the Fourier integral~\eqref{Psea}, it is clear from
Lorentz symmetry that for every~$x$ and~$y$ for which the Fourier integral
exists, $P(x,y)$ can be written as
\beq \label{Pxyrep}
P(x,y) \;=\; \alpha\, (y-x)_j \gamma^j + \beta\:\1
\eeq
with two complex coefficients~$\alpha$ and~$\beta$. Taking the conjugate, we see that
\[ P(y,x) \;=\; \overline{\alpha}\, (y-x)_j \gamma^j + \overline{\beta}\:\1 \:. \]
As a consequence,
\beq \label{1}
A_{xy} \;=\; P(x,y)\, P(y,x) \;=\; a\, (y-x)_j \gamma^j + b\, \1
\eeq
with real parameters~$a$ and $b$ given by
\beq \label{ab}
a \;=\; \alpha \overline{\beta} + \beta \overline{\alpha} \:,\spc
b \;=\; |\alpha|^2 \,(y-x)^2 + |\beta|^2 \:.
\eeq
Applying the formula~$(A_{xy} - b \1)^2 = a^2\:(y-x)^2$, one can easily compute the roots
of the characteristic polynomial of~$A_{xy}$,
\beq \label{lambdaj}
\lambda_1 \;=\; \lambda_2 \;=\; b + \sqrt{a^2\: (y-x)^2} \:,\spc
\lambda_3 \;=\; \lambda_4 \;=\; b - \sqrt{a^2\: (y-x)^2}\:.
\eeq
If the vector~$(y-x)$ is timelike, we see from the inequality~$(y-x)^2>0$ that
the~$\lambda_j$ are all real. If conversely the vector~$(y-x)$ is spacelike,
the term~$(y-x)^2<0$ is negative. As a consequence, the~$\lambda_j$ form a complex conjugate
pair. This shows that for Dirac spinors in Minkowski space,
Definition~\ref{defcausal} is indeed consistent with the usual notion of causality.
Moreover, from the fermionic projector~\eqref{Psea} one can deduce the Minkowski metric as well as all
all the familiar objects of Dirac theory. For example,
the non-negative quantity~$\Sl \psi | \gamma^0 \psi \Sr$ has the interpretation as the
probability density of the Dirac particle. Polarizing and integrating over space yields the scalar product
\begin{equation} \label{print}
(\psi | \phi) = \int_{t=\text{const}} \Sl \psi(t,\vec{x}) \,|\, \gamma^0 \phi(t,\vec{x}) \Sr \:d \vec{x}\:.
\end{equation}
For solutions of the Dirac equation, current conservation implies that this 
scalar product is time independent. In this way, we can describe Dirac theory in Minkowski space
in the setting of fermionic projector.

A more difficult question is if and in which sense the distribution~\eqref{Psea}
is a minimizer of our action principle~\eqref{action}. We refer the interested reader to
the survey article~\cite{lrev} and the references therein.

\subsection{A Formulation of Quantum Field Theory} \label{seccontinuum}
The above correspondence to vacuum Dirac sea structures can also be used to analyze our
action principle for interacting systems in the so-called {\em{continuum limit}}.
We now outline a few ideas and constructions
(for details see~\cite[Chapters~4]{PFP}, \cite{sector} and the survey article~\cite{srev}).
First, it is helpful to observe that the vacuum fermionic projector~\eqref{Psea}
is a solution of the Dirac equation~$(i \gamma^j \partial_j - m) P^\text{sea}(x,y)=0$.
To introduce the interaction, we replace the free Dirac operator by a more general Dirac operator,
for example involving gauge potentials or a gravitational field. Thus, restricting attention
to the simplest case of an electromagnetic potential~$A$, we demand that
\beq \label{DiracP}
\left( i \gamma^j (\partial_j - ie A_j) - m \right) P(x,y) = 0 \:.
\eeq
Moreover, we introduce particles and anti-particles by occupying (suitably normalized)
positive-energy states and removing states of the sea,
\beq \label{particles}
P(x,y) = P^\text{sea}(x,y)
-\frac{1}{2 \pi} \sum_{k=1}^{n_f} |\psi_k(x) \Sr \Sl \psi_k(y) |
+\frac{1}{2 \pi} \sum_{l=1}^{n_a} |\phi_l(x) \Sr \Sl \phi_l(y) | \:.
\eeq
Using the so-called causal perturbation expansion and light-cone expansion,
the fermio\-nic projector can be introduced uniquely from~\eqref{DiracP} and~\eqref{particles}.

It is important that our setting so far does not involve the field equations; in particular,
the electromagnetic potential in the Dirac equation~\eqref{DiracP} does not need to satisfy
the Maxwell equations. Instead, the field equations should be derived from
our action principle~\eqref{action}. Indeed, analyzing the corresponding Euler-Lagrange equations,
one finds that they are satisfied only if the potentials in the Dirac equation satisfy certain
constraints. Some of these constraints are partial differential equations involving
the potentials as well as the wave functions of the particles and anti-particles in~\eqref{particles}.
In~\cite{sector}, such field equations are analyzed in detail for a system involving an
axial field. In order to keep the setting as simple as possible, we here consider the analogous
field equation for the electromagnetic field
\beq \label{Maxwell}
\partial_{jk} A^k - \Box A_j = e \sum_{k=1}^{n_f} \Sl \psi_k | \gamma_j \psi_k \Sr
-e \sum_{l=1}^{n_a} \Sl \phi_l | \gamma_j \phi_l \Sr \:.
\eeq
With~\eqref{DiracP} and~\eqref{Maxwell}, the interaction as described by the
action principle~\eqref{action} reduces in the continuum limit to the
{\em{coupled Dirac-Maxwell equations}}.
The many-fermion state is again described by the fermionic projector, which is built up
of {\em{one-particle wave functions}}. The electromagnetic field merely is a
{\em{classical bosonic field}}. Nevertheless, regarding~\eqref{DiracP} and~\eqref{Maxwell}
as a nonlinear hyperbolic system of partial differential equations and treating
it perturbatively, one obtains all the Feynman diagrams which do not involve fermion loops.
Taking into account that by exciting sea states we can describe pair creation and annihilation
processes, we also get all diagrams involving fermion loops.
In this way, we obtain agreement with perturbative quantum field theory
(for details see~\cite[\S8.4]{sector}).

\section{Microscopic Mixing of Decoherent Subsystems} \label{secmix}
In the recent paper~\cite{entangle}, a method was developed for describing entanglement
and second quantized fermionic and bosonic fields in the framework of the fermionic projector.
We now outline a few constructions and results from this paper.

\subsection{Description of Entangled Second Quantized Fermionic Fields} \label{secentangle}
Recall that in the setting of Section~\ref{secdiscrete} we described the fermion system
by a projector~$P$ on an~$f$-dimensional, negative definite subspace of~$H$.
It is now useful to choose an orthonormal basis~$\psi_1, \ldots, \psi_f$ of~$P(H)$
(meaning that~$\bra \psi_i | \psi_j \ket = - \delta_{ij}$).
Then the discrete kernel of the fermionic projector can be written in bra/ket-notation as
\beq \label{Pdiscrete}
P(x,y) = -\sum_{j=1}^f |\psi_j(x) \Sr \Sl \psi_j(y) |\:.
\eeq
In order to get a connection to the Fock space formalism, we take the wedge product
of the one-particle wave functions,
\beq  \label{wedge}
\Psi := \psi_1 \wedge \cdots \wedge \psi_f \:,
\eeq
to obtain a vector in the fermionic Fock space.
This shows that the fermions can indeed be described by a second-quantized field. However, as~$\Psi$
is a Hartree-Fock state, it is impossible to describe entanglement
(see for example~\cite[Example~3.1]{entangle}).

This shortcoming is removed by introducing the concept of {\em{microscopic mixing}}.
It is based on the assumption is that~$P$ has a non-trivial microstructure.
This assumption seems natural in view of setting in discrete space-time (see Section~\ref{secdiscrete}),
where space-time is not smooth on the microscopic scale.
``Averaging'' this microstructure over macroscopic regions of space-time gives rise to an effective
kernel~$P(x,y)$ of a more general form, making it possible to describe entanglement.

\begin{figure}
\begin{center}
\begin{picture}(0,0)%
\includegraphics{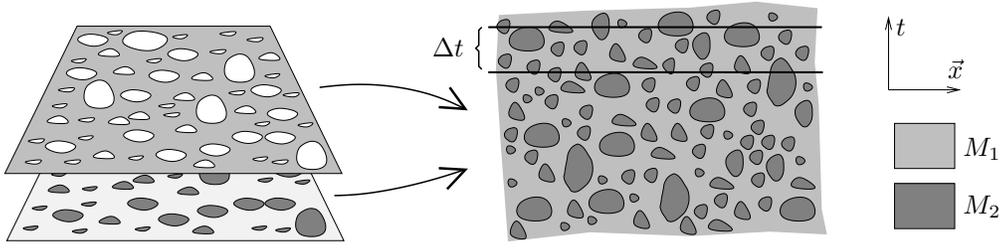}%
\end{picture}%
\setlength{\unitlength}{829sp}%
\begingroup\makeatletter\ifx\SetFigFontNFSS\undefined%
\gdef\SetFigFontNFSS#1#2#3#4#5{%
  \reset@font\fontsize{#1}{#2pt}%
  \fontfamily{#3}\fontseries{#4}\fontshape{#5}%
  \selectfont}%
\fi\endgroup%
\begin{picture}(29775,7182)(-1319,903)
%\put(-554,5654){\makebox(0,0)[b]{\smash{{\SetFigFontNFSS{10}{12.0}{\rmdefault}{\mddefault}{\updefault}$M^\text{cont}_1$}}}}
%\put(-1304,1844){\makebox(0,0)[b]{\smash{{\SetFigFontNFSS{10}{12.0}{\rmdefault}{\mddefault}{\updefault}$M^\text{cont}_2$}}}}
\put(12481,6449){\makebox(0,0)[b]{\smash{{\SetFigFontNFSS{10}{12.0}{\rmdefault}{\mddefault}{\updefault}$\Delta t$}}}}
\put(28441,3449){\makebox(0,0)[b]{\smash{{\SetFigFontNFSS{10}{12.0}{\rmdefault}{\mddefault}{\updefault}$M_1$}}}}
\put(28441,1739){\makebox(0,0)[b]{\smash{{\SetFigFontNFSS{10}{12.0}{\rmdefault}{\mddefault}{\updefault}$M_2$}}}}
\put(27646,5774){\makebox(0,0)[b]{\smash{{\SetFigFontNFSS{10}{12.0}{\rmdefault}{\mddefault}{\updefault}$\vec{x}$}}}}
\put(26026,7064){\makebox(0,0)[b]{\smash{{\SetFigFontNFSS{10}{12.0}{\rmdefault}{\mddefault}{\updefault}$t$}}}}
\end{picture}%
\caption{Example of a microscopic mixing of two space-time regions.}
\label{fig2}
\end{center}
\end{figure}

To explain the construction, suppose that we want to describe
the fermionic state $\Psi = \Psi^{(1)} + \Psi^{(2)}$, being a superposition of two
states, each of which is an $f$-particle Hartree-Fock state of the form
\beq \label{wedgea}
\Psi^{(a)} := \phi_1^{(a)} \wedge \cdots \wedge \phi_f^{(a)} \qquad
(a=1,2)
\eeq
(the generalization to an arbitrary finite number of subsystems is straightforward).
We now decompose Minkowski space~$M$ into two disjoint subsets
\beq \label{subs}
M = M_1 \cup M_2 \qquad \text{with} \qquad M_1 \cap M_2 = \varnothing \:,
\eeq
which should be fine-grained in the sense that every macroscopic region of space-time intersects
both subsets (as indicated in Figure~\ref{fig2}).
We introduce one-particle wave functions~$\psi_j$ which on~$M_a$ coincide with~$\phi_j^{(a)}$,
\beq \label{psisplit}
\psi_j = \psi_j^{(1)} + \psi_j^{(2)} \qquad \text{with} \qquad
\psi_j^{(a)} = \chi_{M_a} \,\phi_j^{(a)}
\eeq
(here~$\chi_{M_a}$ denotes the characteristic function defined by
$\chi_{M_a}(x)=1$ if~$x \in M_a$ and $\chi_{M_a}(x)=0$ otherwise).
The form of the resulting fermionic projector~\eqref{Pdiscrete} depends
on whether its arguments~$x$ and~$y$ are in~$M_1$ or~$M_2$,
\[ P(x,y) = -\sum_{j=1}^f |\psi_j^{(a)}(x) \Sr \Sl \psi_j^{(b)}(y) |
\quad \text{if~$x \in M_a$ and~$y \in M_b$} \:. \]
If both arguments are in the same subsystem, then the fermionic projector is composed
of the wave functions of this subsystem. But if the arguments are in different subsystems,
then the fermionic projector involves both wave functions~$\phi_j^{(1)}$
and~$\phi_j^{(2)}$, thus giving correlations between the two subsystems. 
The idea is to remove all these correlation terms by bringing
the wave functions of the subsystems ``out of phase.'' In technical terms, we transform the wave functions 
of the second subsystem by a unitary matrix~$U$,
\beq \label{decoher}
\psi_j^{(1)} \rightarrow \psi_j^{(1)} \:,\qquad
\psi_j^{(2)} \rightarrow \tilde{\psi}_j^{(2)} :=
\sum_{k=1}^f U_{jk} \,\psi_k^{(2)} \quad \text{with} \qquad U \in \SU(f)\:.
\eeq
This transformation does not change the fermionic projector if its two arguments are
in the same subsystem, because in the case~$x,y \in M_2$, the unitarity of~$U$ yields that
\begin{align*}
P(x,y) \rightarrow &-\sum_{j=1}^f  |\tilde{\psi}^{(2)}_j(x) \Sr \Sl \tilde{\psi}^{(2)}_j(y) |
= -\sum_{j,k=1}^f (U U^\dagger)_{jk} \:| \psi^{(2)}_j(x) \Sr \Sl \psi^{(2)}_k(y) | \\
&= -\sum_{j}^f |\psi^{(2)}_j(x) \Sr \Sl \psi^{(2)}_j(y) | = P(x,y) \:.
\end{align*}
However, if the two arguments of the fermionic projector are in different space-time regions,
the operator~$U$ does not drop out. For example, if~$x \in M_2$ and~$y \in M_1$,
\beq \label{Psum}
P(x,y) \rightarrow -\sum_{j=1}^f |\tilde{\psi}^{(2)}_j(x) \Sr \Sl \psi^{(1)}_j(y) |
= -\sum_{j,k=1}^f U_{jk} \:| \psi^{(2)}_j(x) \Sr \Sl \psi^{(1)}_k(y) | \:.
\eeq
In the special case when~$U$ is a diagonal matrix whose entries are phase factors,
\[ U = \text{diag}(e^{i \varphi_1}, \ldots, e^{i \varphi_f}) \qquad \text{with} \qquad
\sum_{j=1}^f \varphi_j = 0 \mod 2 \pi \:, \]
the summations in~\eqref{Psum} reduce to one sum involving the phase factors,
\[ P(x,y) \rightarrow -\sum_{j=1}^f e^{i \varphi_j} \:|\psi^{(2)}_j(x) \Sr \Sl \psi^{(1)}_j(y) | \:. \]
If the angles~$\varphi_j$ are chosen stochastically, the phases of the summands are random.
As a consequence, there will be cancellations in the sum.
At this point, it is important to remember that the number~$f$ of fermions of our system
is very large, because we also count the states of the Dirac sea
(see Section~\ref{seccontinuum}). As a consequence, $P(x,y)$ will be very small.
More generally, we find that if~$U$ is a random matrix,
$P(x,y)$ involes a scale factor~$f^{-\frac{1}{2}}$ if~$x$ and~$y$ lie in different subsystems.

Next we need to homogenize the wave functions by ``taking averages'' over macroscopic
regions of space-time. Our concept is that this homogenization should take place in the
quantum mechanical measurement process, because the measurement device is also composed
of wave functions which are ``smeared out'' on the microscopic scale. In order to implement this
idea, we introduce the so-called {\em{measurement scalar product}}, where we replace
the spatial integral in~\eqref{print} by integrals over a strip of width~$\Delta t$ in space-time
(see Figure~\ref{fig2} and the more detailed treatment in~\cite[Section~4.2 and~4.3]{entangle}).
As a consequence of the homogenization process, the system is described effectively by the sum of
the fermionic projectors corresponding to the two subsystems.

In order to clarify the above construction in the Fock space formalism, we next consider the
many-particle wave function~\eqref{wedge}. Using~\eqref{psisplit}, we obtain
\beq \label{multip}
\Psi = \psi_1 \wedge \cdots \wedge \psi_f =
(\psi_1^{(1)} + {\psi}_1^{(2)}) \wedge \cdots \wedge (\psi_f + {\psi}_f)\:.
\eeq
Multiplying out, we obtain many summands. Two of these summand correspond to
the Hartree-Fock states of the two subsystems~\eqref{wedgea},
but we obtain many other contributions. These additional contributions are again removed
by the transformation~\eqref{decoher}. Namely, this transformation leaves the
many-particle wave functions of the subsystems remain unchanged, because
\beq \label{Psi2}
\Psi^{(2)} \rightarrow \tilde{\Psi}^{(2)} = \tilde{\psi}_1^{(2)} \wedge \cdots \wedge \tilde{\psi}_f^{(2)}
= \det U \; \psi_1^{(2)} \wedge \cdots \wedge \psi_f^{(2)} = \Psi^{(2)} \:.
\eeq
But all the summands obtained by multiplying out~\eqref{multip} which
involve factors~$\psi_j^{(1)}$ as well as~$\psi_k^{(2)}$ contain matrix elements of~$U$.
These matrix elements again become small if~$f$ gets large. As a consequence,
the system is described effectively by the sum of the Hartree-Fock states of the subsystems.
The homogenization process can again be described by the measurement scalar product.
It is most convenient to introduce on the many-particle wave functions the scalar product induced
by the measurement process, giving rise to the effective fermionic Fock space~$\F^\text{eff}$.

The above consideration can be understood using the notion of decoherence. If the one-particle
wave functions~$\psi_j^{(1)}$
and~$\psi_j^{(2)}$ are coherent or ``in phase'', then the fermionic projector~$P(x,y)$
has the usual form, no matter whether~$x$ and~$y$ are in the same subsystem or not.
If however the wave functions in the subregions are decoherent or ``out of phase'', then the
fermionic projector~$P(x,y)$ becomes very small if~$x$ and~$y$ are in different subregion.
We refer to this effect as the {\em{decoherence between space-time regions}}.
It should be carefully distinguished from the decoherence of the many-particle wave function
(see for example~\cite{joos}). Namely, as we saw in~\eqref{Psi2}, in our case
the many-particle wave functions remain unchanged. Thus they remain coherent, no
decoherence in the sense of~\cite{joos} appears. But the one-particle wave functions
are decoherent~\eqref{decoher}, implying that correlations between the two subsystems
become small.

We finally remark that the mixing of subsystems can be generalized to the so-called
{\em{holographic mixing}}, where the subsystems need no longer be localized in disjoint
regions of space-time. We refer the interested reader to~\cite[Section~5.3]{entangle}.

\subsection{Description of Second Quantized Bosonic Fields} \label{secSQ}
As outlined in Section~\ref{seccontinuum}, in the continuum limit the interaction is described
by the Dirac equation coupled to classical field equations. If our system involves a
microscopic mixing of several subsystems, we can take the continuum limit
of each subsystem separately. Labeling the subsystems by an index~$a = 1,\ldots, L$
and again considering for simplicity the electromagnetic interaction, the
interaction of the $a^\text{th}$ subsystem is described in analogy to~\eqref{DiracP}
and~\eqref{Maxwell} by the coupled Dirac-Maxwell equations
\beq \begin{split} \label{DMsub}
\Big( i \gamma^j (&\partial_j - ie A^{(a)}_j) - m \Big) P^{(a)}(x,y) = 0 \\
\partial_j^{\;\,k} A^{(a)}_k - \Box A^{(a)}_j &= e \sum_{k=1}^{n_f} \Sl \psi^{(a)}_k | \gamma_j \psi^{(a)}_k \Sr
-e \sum_{l=1}^{n_a} \Sl \phi^{(a)}_l | \gamma_j \phi^{(a)}_l \Sr\:.
\end{split}
\eeq
In particular, the subsystems have an independent dynamics as described by these equations.
To clarify the physical picture, we again point out that the underlying space-time is not smooth
on the microscopic scale, but it is decomposed into subsystems~$M_a$ which are
microscopically mixed.
If~$x$ and~$y$ are in different subsystems, the fermionic projector~$P(x,y)$ is very small
due to decoherence, implying that the causal relations of Definition~\ref{defcausal} no longer
agree with the causal structure in Minkowski space. In other words, the usual causal relations
are not valid between points in different subsystems. This means that the subsystems should be
regarded as having a simultaneous and independent existence in our space-time
(as indicated in Figure~\ref{fig2} by the two ``space-time sheets'').

The bosonic field in~\eqref{DMsub} is purely classical. The key observation for
passing from a classical to a second-quantized bosonic field is that the dynamics of a free
second-quantized bosonic field can be described equivalently by a superposition of solutions
of the classical field equations (for details see~\cite[Section~5.1]{entangle}).
Thus to describe a free second-quantized bosonic field, it suffices to consider
the microscopic mixing of subsystems~\eqref{DMsub} (each with a classical
bosonic field~$A^{(a)}$ satisfying the classical field equations),
and to associate to every subsystem a complex number~$\phi(a)$.
The resulting function~$\phi(a)$ can be interpreted as a wave function
on the space of classical field configurations. For clarity, we first construct~$\phi(a)$ in the
case when no fermions are present. In this case, we take the wedge product of all the states~$\psi^{(a)}_j$
which form the fermionic projector of the subsystem~$a$ to obtain a many-particle wave
function~$\Psi^{(a)}$.
We then compare this wave function with the corresponding wave function~$\check{\Psi}$ of a reference system (i.e.\ of a Dirac sea configuration in the presence of the external field~$A^{(a)}$). These two
wave functions coincide up to a proportionality factor, and thus we can define~$\phi(a)$ by
\beq \label{Psia}
\Psi^{(a)} = \phi(a)\: \check{\Psi} \qquad \text{with} \qquad \phi(a) \in \C\:.
\eeq

The last construction immediately generalizes to the situation when fermionic particles and/or anti-particles
are present. In this case, one decomposes the many-particle state into a component
involving the particles and anti-particle states in~\eqref{particles} and the sea sates,
\beq \label{Psifermbos}
\Psi^{(a)} = \Big( \psi_1^{(a)} \wedge \cdots \wedge \psi_{n_f}^{(a)}
\wedge \phi_1^{(a)} \wedge \cdots \wedge \phi_{n_a}^{(a)} \Big)
\wedge \Big[ \psi_{n+1} \wedge \cdots \wedge \psi_f \Big] \:.
\eeq
Then after homogenizing, the particles and anti-particles give rise to a vector of the
fermionic Fock space~$\F^\text{eff}$, whereas the sea states can be used to define in
analogy to~\eqref{Psia} the second-quantized bosonic wave function~$\phi$.

We finally point that in the above constructions, we worked with {\em{free}} bosonic fields.
To get the connection to interacting quantum fields, we recall that in the continuum limit
(see Section~\ref{seccontinuum} above), one gets all the Feynman diagrams.
We thus obtain agreement also with the quantitative predictions of perturbative quantum field theory.
Nevertheless, the fermionic projector approach is not equivalent to a local
interacting quantum field theory (for example in the canonical formulation on a Fock space).
Clarifying the connections and differences of the two frameworks is work in progress.

\section{Physical Interpretation} \label{secinterpret}
We now discuss the previous constructions and results from a conceptual point of view.
For clarity, we try to explain the physical picture in simple examples,
without aiming for mathematical rigor nor maximal generality.

\subsection{The Superposition Principle} \label{secsuper}
In the standard formulation of quantum physics, it is a general principle
that superpositions of quantum states can be formed (for a good exposition see~\cite[Section~2.1]{joos}).
The superposition principle holds for the one-particle wave functions in quantum mechanics just
as well as for the quantum state describing the whole physical system.
In the framework of the fermionic projector, however, where the whole system is described by a projector
in an indefinite inner product space, the validity of the superposition principle is not
obvious. We now explain this point in detail.

In the framework of the fermionic projector, the superposition principle arises on
different levels. On the fundamental level of discrete space-time, the wave functions
are vectors in the indefinite inner product space~$(\H, \bra .|. \ket)$, so that
superpositions of one-particle wave functions can be formed.
In the continuum limit, where the interaction is described by the Dirac 
equation~\eqref{DiracP} coupled to a classical field~\eqref{Maxwell},
we thus obtain the superposition principle for the Dirac wave functions. Moreover, since the
Maxwell equations are linear, the superposition principle also holds for
classical electromagnetic waves.

For many-particle states, however, the superposition principle does not hold on the fundamental level.
In particular, it is impossible to form the naive superposition of two physical systems,
simply because the linear combination of two projectors in general is no longer a projector.
But the superposition principle does hold for the effective fermionic many-particle wave function
obtained by decomposing the system into decoherent subsystems and
homogenizing on the microscopic scale (see Section~\ref{secentangle}).
For the free second-quantized electromagnetic field as described in Section~\ref{secSQ},
the superposition principle corresponds to taking linear combinations of the
complex-valued wave function~$\phi$ defined on the classical field configurations.
Since the function~$\phi$ is constructed out of the fermionic many-particle wave function
(see~\eqref{Psia}),
linear combinations are again justified exactly as for the fermionic many-particle.

We conclude that in the framework of the fermionic projector, the superposition principle
again holds. It is possible to form linear combinations of macroscopic systems (like a
dead and a living cat). Nevertheless, the framework of the fermionic projector
differs from standard quantum theory in that for many-particle wave functions, the superposition
principle does not hold on the fundamental level, but it is merely a consequence of the
microscopic mixing of decoherent subsystems.
This means that the superposition principle is overruled in situations when
our action principle in discrete space-time~\eqref{action} needs to analyzed
beyond the continuum limit. We will come back to this point 
in Section~\ref{seccollapse} in the context of collapse phenomena.

\subsection{The Measurement Problem and Decoherence} \label{secmeasure}
One of the most controversial and difficult points in the understanding of quantum physics
is the so-called measurement problem. In simple terms, it can be understood from the
dilemma that on one side, the dynamics of a quantum system is described by a linear evolution
equation in a Hilbert space (for example the Schr\"odinger equation), and considering the
measurement apparatus as part of the system, one would expect that this linear and
deterministic quantum evolution alone should give a complete description of physics.
But on the other side, the Copenhagen interpretation requires an external observer, who
by making a measurement triggers a ``collapse'' or ``reduction'' of the wave function to an
eigenstate of the observable. It is not obvious how the external observer can be described
within the linear quantum evolution. Also, the statistical interpretation of the expectation value
in the measurement process does not seem to correspond to the deterministic nature of the
quantum evolution.

The measurement problem has been studied extensively in the literature, and many different
solutions have been proposed (see for example~\cite{wheeler+zurek, duerr+teufel, joos, cini,
albert, pearle}).
Here we shall not try to enter an exhaustive discussion or comparison of the different interpretations
of quantum mechanics. We only explain how our concepts fit into the picture and give
one possible interpretation which corresponds to the personal preference of the author.
But it is well possible that the framework of the fermionic projector can be adapted to other
interpretations as well.

\begin{figure}
\begin{center}
\begin{picture}(0,0)%
\includegraphics{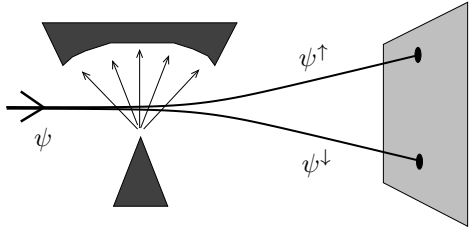}%
\end{picture}%
\setlength{\unitlength}{829sp}%
\begingroup\makeatletter\ifx\SetFigFontNFSS\undefined%
\gdef\SetFigFontNFSS#1#2#3#4#5{%
  \reset@font\fontsize{#1}{#2pt}%
  \fontfamily{#3}\fontseries{#4}\fontshape{#5}%
  \selectfont}%
\fi\endgroup%
\begin{picture}(13891,6749)(6922,637)
\put(16186,5444){\makebox(0,0)[b]{\smash{{\SetFigFontNFSS{10}{12.0}{\rmdefault}{\mddefault}{\updefault}$\psiu$}}}}
\put(16276,2294){\makebox(0,0)[b]{\smash{{\SetFigFontNFSS{10}{12.0}{\rmdefault}{\mddefault}{\updefault}$\psid$}}}}
\put(8086,3044){\makebox(0,0)[b]{\smash{{\SetFigFontNFSS{10}{12.0}{\rmdefault}{\mddefault}{\updefault}$\psi$}}}}
\end{picture}%
\caption{The Stern-Gerlach experiment.}
\label{figsg}
\end{center}
\end{figure}

We begin by considering the {\em{Stern-Gerlach experiment}}. Thus a beam of atoms
passes through an inhomogeneous magnetic field.
Decomposing the wave function~$\psi$ into the components with spin up and down,
\beq \label{psiud}
\psi = \psiu + \psid \:,
\eeq
these two components feel opposite magnetic forces. As a consequence, the beam splits
up into two beams, leading to two exposed dots on the photographic material (see Figure~\ref{figsg}).
If the intensity of the beam is so low that only one atom passes through the magnetic field, then
either the upper or the lower dot will be exposed,
both with probability one half. It is impossible to predict whether the electron will fly up
or down; only probabilistic statements can be made.

Let us try to describe the Stern-Gerlach experiment in the framework of the fermionic projector.
For simplicity, we replace the atom by an electron (disregarding the
Lorentz force due to the electron's electric charge). Then at
the beginning, the system is described by a classical external magnetic field and a
Dirac wave function~$\psi(t, \vec{x})$ which at time~$t=0$ has the form of
a wave packet moving towards the magnetic field. This situation is modeled
by the fermionic projector in the continuum limit~\eqref{particles}
for one particle; the dynamics is described by the Dirac equation~\eqref{DiracP} in the given external
field. Solving the Dirac equation, the wave function splits into two components~\eqref{psiud},
which are deflected upwards and downwards, respectively.
Writing the contribution of the wave function to the fermionic projector~\eqref{particles} as
\begin{align}
-\frac{1}{2 \pi}\: |\psi(x) \Sr \Sl \psi(y)| =  -\frac{1}{2 \pi} \Big( &
 |\psiu(x) \Sr \Sl \psiu(y)| +  |\psid(x) \Sr \Sl \psid(y)| \label{psid} \\
& +  |\psiu(x) \Sr \Sl \psid(y)| +  |\psid(x) \Sr \Sl \psiu(y)| \Big) , \label{psiad}
\end{align}
one gets contributions of different type. Namely, the two summands in~\eqref{psid} are localized
at the upper and lower electron beam, respectively. The two summands in~\eqref{psiad}, however, are delocalized and give correlations between the two beams.
As observed in~\cite[Chapter~10]{sector}, the Euler-Lagrange equations in the continuum limit
cannot be satisfied if general delocalized contributions to the fermionic projector are present.
This means that there should be a mechanism which tries to avoid nonlocal
correlations as in~\eqref{psiad}. A possible method for removing the nonlocal
correlations is to divide the system into two decoherent subsystems 
(as shown in Figure~\ref{fig2}, although at this stage they do not necessarily need to be
microscopically mixed), in such a way that~$\psiu$ belongs to the first and~$\psid$ to the
the second subsystem. Then the continuum limit is to be taken separately in the two
subsystems. The contribution of the wave function in the two subsystems simply is
\[ -\frac{1}{2 \pi} \: |\psiu(x) \Sr \Sl \psiu(y)| 
\qquad \text{and} \qquad
-\frac{1}{2 \pi} \: |\psid(x) \Sr \Sl \psid(y)| \:, \]
respectively. Thus the delocalized terms~\eqref{psiad} no longer occur, so that the
problem of solving the Euler-Lagrange equations observed in~\cite[Chapter~10]{sector}
has disappeared. This consideration gives a possible mechanism for the
{\em{generation of subsystems}}.

Let us carefully discuss different notions of {\em{decoherence}}. First of all, the space-time
points of the two subsystems should be decoherent, in the sense that the states of the
fermionic projector are unitarily transformed in the second subsystem~\eqref{decoher}.
This decoherence has the effect that the fermionic projector~$P(x,y)$ becomes very small if~$x$
and~$y$ are in different subsystems (see~\eqref{Psum}), implying that the two subsystems have an independent dynamics in the continuum limit.
However, the many-particle wave function of the system is not
affected by the decoherence between the space-time points (see~\eqref{Psi2}).
Rewriting it similar to~\eqref{Psifermbos} as the wedge product of the one-particle wave
function~$\psi$ with the sea states, one sees that the quantum mechanical wave functions
of the two subsystems are still coherent. In particular, if the two beams interfered with each other
(for example after redirecting them with additional Stern-Gerlach magnets), they could be
superposed quantum-mechanically, giving rise to the usual interference effects
of the double slit experiment.
We also point out that the dynamics of each subsystem is still described by the Dirac
equation~\eqref{DiracP} in the external magnetic field. Since the Dirac equation is linear,
solving it for~$\psi$ is the same as solving it separately for the two components~$\psiu$
and~$\psid$. Thus at this point, the dynamics is not affected by the decomposition into
subsystems; we still have the linear deterministic dynamics as described by the Dirac equation.

As just explained, at this stage the generation of subsystems has no effect on the dynamics
of the system. This suggests that it should not be observable whether the subsystems have
formed or not. This motivates us to demand that expectation values taken with respect to the
measurement scalar product (as introduced in Section~\ref{secentangle}) should not be affected
by the generation of subsystems. Keeping in mind that, after a suitable homogenization process,
the measurement scalar product coincides with the integral~\eqref{print},
we can say alternatively that the process of {\em{generation of subsystems should preserve the
probability densities}}. This condition also ensures that when we get the
connection to the statistical description of the measurement process, the
probabilities are indeed given by the spatial integrals of the absolute square of the wave functions.
Since the probability density is the zero component of the probability current~$\Sl \psi | \gamma^j | \psi \Sr$,
we can say equivalently that the generation of subsystems should respect {\em{current conservation}}.
This assumption seems reasonable, because current conservation holds in each subsystem
as a consequence of the Dirac equation, and we merely extend this conservation law to
the situation when the number of subsystems changes.

The dynamics becomes more complicated when the wave function approaches the
screen, because the interaction with the electrons of the photographic material can no longer be
described by an external field. Instead, one must consider the coupled Dirac-Maxwell
equations~\eqref{DiracP} and~\eqref{Maxwell} for a many-electron system.
Since the interaction is no longer linear, it now makes a difference that
the two subsystems have an independent dynamics. More specifically, in the first subsystem
the wave function in the upper beam interacts with the electrons near the upper impact
point, whereas the second subsystem describes the interaction of the lower beam.
The whole system is a superposition of these two systems, described mathematically
by a vector in the Fock space~$\F^\text{eff}$.
Exactly as explained in~\cite[Chapter~3]{joos}, the different interaction with the environment
leads to a decoherence of the many-particle wave functions of the two subsystems.
Thus now it is no longer possible to form quantum mechanical superpositions of the
wave functions of the two subsystems. The whole system behaves like a
statistical ensemble of the two subsystems.
Following the resolution of the measurement problem as proposed in~\cite{joos},
one should regard the human observer as being part of the system.
Thus the observer is also decomposed into two observers, one in each subsystem.
The two observers measure different outcomes of the experiment.
Due to the decoherence of their wave functions, the two observers cannot communicate
with each other and do not even experience the existence of their counterparts.
From the point of view of the observer, the outcome of the experiment can only be described
statistically: the electron moves either up or down, both with probability one half.

Other experiments like the {\em{spin correlation experiment}} can be understood similarly. One only
needs to keep in mind that if entanglement is present, then the subsystems must be microscopically mixed.
The quantum state of each subsystem is not entangled. But homogenizing on the microscopic scale 
leads to an effective description of the system by an entangled state in the
Fock space~$\F^\text{eff}$.

We finally remark that the mechanism for the generation of subsystems proposed above
could be made mathematically precise by analyzing the action principle~\eqref{action}
in the discrete setting, going beyond the approximation of the continuum limit.
One should keep in mind that on this level, our action principle {\em{violates causality}}. Thus it
is conceivable that the formation of subsystems depends on later measurements or that
subsystems tend to form eigenstates of the subsequent measurement device.
However, such effects can hardly be verified or falsified in experiments, and therefore we
will not consider them here (for a discussion of a measurable effect of causality
violation see~\cite[Section~8.2]{sector}).

\subsection{The Wave-Particle Duality} \label{secwpd}
Following the above arguments, one can also understand the wave-particle duality in a
way where the wave function is the basic physical object, whereas the particle character
is a consequence of the interaction as described by the action principle~\eqref{action}.
To explain the idea, we return to the Stern-Gerlach experiment of Figure~\ref{figsg}.
In the previous section, we justified that if one electron flies through the magnetic field,
the atom will expose either the upper or the lower dot on the screen, much in contrast to
the behavior of a classical wave, which would be observable at both dots of the
screen at the same time.
Repeating the arguments of the previous section on the scale of the atoms of the
photographic material, we conclude that more and more subsystems will form,
which become decoherent as explained in~\cite[Chapter~3]{joos}.
For an observer in one of the subsystems, the electron will not expose
the whole dot on the screen uniformly, but it will only excite one atom of the photographic
material. As a consequence, the electron appears like a point particle.
Again, the outcome of the measurement can only be described statistically.

\subsection{The Collapse of the Wave Function} \label{seccollapse}
The resolution of the measurement problem in Section~\ref{secmeasure}
is conceptually convincing and explains
the experimental observations. It goes back to Everett's ``relative state interpretation''
and has found many different variations (see~\cite{wheeler+zurek} and~\cite{cini}).
But no matter which interpretation one prefers, there always remains the counter-intuitive effect
that all the possible outcomes of experiments are realized as components of the quantum state of
the system. Thus when time evolves, the quantum state disintegrates into more and more
decoherent components, which should all describe a physical reality.
This phenomenon, which is often subsumed under the catchy but oversimplified title
``many-worlds interpretation,'' is difficult to imagine and hard to accept.
Another criticism is that decoherence leads to an effective description by
a density operator, which however does not uniquely determine the Fock states of the
decoherent components (for details see~\cite[Section~4]{ghirardi} or~\cite[Section~6]{albert}).
In order to avoid these problems, it has been proposed to introduce a
mechanism which leads to a ``collapse'' or ``reduction'' of the wave function.
Different mechanism for a collapse have been discussed, in particular
models where the collapse occurs at discrete time steps~\cite{ghirardi1}
or is a consequence of a stochastic process~\cite{pearle0}.
These models have in common that the superposition principle is overruled
by a nonlinear component in the dynamics. The nonlinearity is chosen to be so weak
that it does not contradict the experimental evidence for a linear dynamics.
In other words, the nonlinear term is so small that it cannot be detected experimentally.
But nevertheless, this term can be arranged to prevent superpositions of macroscopically
different wave functions.

In the framework of the fermionic projector, the dynamics as described by the
action principle~\eqref{action} is nonlinear. This indeed
provides a {\em{new collapse mechanism}},
as we now explain. Suppose that our system is described by many decoherent
subsystems. Since perfect decoherence seems impossible to arrange,
the kernel of the fermionic projector~$P(x,y)$ will in general not vanish identically
if~$x$ and~$y$ are in two different subsystems. This gives rise to a contribution
to the action~\eqref{action} which, although being small due to decoherence, 
is strictly positive. These contributions, which ``mix'' different subsystems, grow
quadratically with the number of subsystems, thus penalizing a very large number of subsystems.
This shows that there is a mechanism which tries to reduce the
number of subsystems, violating the superposition principle and the independent dynamics
of the subsystems. It seems very plausible that this mechanism leads to a collapse of the wave function
(although a derivation from the  action principle~\eqref{action} or a quantitative analysis
has not yet been given). We also point out that, just as in other collapse theories,
the collapse itself seems very difficult to observe. Namely, as systems whose
quantum states are decoherent no longer interact with each other, it is impossible to decide whether there
are other worlds beside the one observed by us, or whether the whole system has collapsed
into the state which we observe. For this reason, despite its importance for the interpretation
of quantum theory, the issue of the collapse of the wave function is often regarded as being speculative.
But at least, the framework of fermion systems in discrete space-time outlined in Section~\ref{secdiscrete}
provides a well-defined mathematical setting for studying collapse phenomena.
Analyzing this framework in more detail might help to overcome the open problems
discussed in~\cite{pearle}, thus leading to a fully satisfying quantum theory. \\[-0.8em]

To summarize the physical interpretation, the framework of the fermionic
projector seems in agreement with the superposition principle and the decoherence
phenomena which explain the appearance of our classical world as well as the wave-particle duality.
Our description is more concrete than the usual Fock space formulation because the
decoherent components of the quantum state should all be realized in space-time by the
states of the fermionic projector.
Moreover, we saw qualitatively that our action principle~\eqref{action} provides a
mechanism for a collapse of the wave function and a reduction of the number of
decoherent components. In view of the fact that this collapse seems very difficult to observe
in experiments, the remaining question of how many ``different worlds'' are realized in our space-time
seems of more philosophical nature. The personal view of the author is that the fermionic
projector should only realize one macroscopic world, but at present this is
mere speculation.

%\bibliographystyle{amsplain}
%\bibliography{../felix}

\def\dbar{\leavevmode\hbox to 0pt{\hskip.2ex \accent"16\hss}d}
\providecommand{\bysame}{\leavevmode\hbox to3em{\hrulefill}\thinspace}
\providecommand{\MR}{\relax\ifhmode\unskip\space\fi MR }
\providecommand{\MRhref}[2]{%
  \href{http://www.ams.org/mathscinet-getitem?mr=#1}{#2}
}
\providecommand{\href}[2]{#2}

\end{document}